# Ultra Low Power Associative Computing with Spin Neurons and Resistive Crossbar Memory

Mrigank Sharad, Deliang Fan, and Kaushik Roy
Department of Electrical and Computer Engineering, Purdue University, West Lafayette, IN, USA
(msharad, dfan, kaushik)@purdue.edu

## ABSTRACT
Emerging resistive-crossbar memory (RCM) technology can be promising for computationally-expensive analog pattern-matching tasks. However, the use of CMOS analog-circuits with RCM would result in large power-consumption and poor scalability, thereby eschewing the benefits of RCM-based computation. We propose the use of low-voltage, fast-switching, magneto-metallic 'spin-neurons' for ultra low-power non-Boolean computing with RCM. We present the design of analog associative memory for face recognition using RCM, where, substituting conventional analog circuits with spin-neurons can achieve ~100x lower power. This makes the proposed design ~1000x more energy-efficient than a 45nm-CMOS digital ASIC, thereby significantly enhancing the prospects of RCM based computational hardware.

## Categories and Subject Descriptors
B.7.1 [Integrated Circuits] Types and Design Styles – Advanced Technologies

## General Terms
Design

## 1. INTRODUCTION

In recent years several device solutions have been proposed for fabricating nano-scale programmable resistive elements, generally categorized under the term 'memristor' [1-9]. Of special interest are those which are amenable to integration with state of the art CMOS technology, like memristors based on Ag-Si filaments [6-8]. Such devices can be integrated into metallic crossbars to obtain high density resistive crossbar memory (RCM) [1-8]. Continuous range of resistance values obtainable in these devices can facilitate the design of multi-level, non-volatile memory [1-3]. The Resistive-Crossbar Memory (RCM) technology has led to interesting possibilities of combining memory with computation [1-5]. RCM can be highly suitable for a class of non-Boolean computing applications that involve pattern-matching [5, 11]. Such applications employ highly memory-intensive computing that may require correlation of a multidimensional input data with a large number of stored patterns or templates, in order to find the best match [11]. Use of conventional digital processing techniques for such tasks incurs high energy and real-estate cost, due to the sheer number of computations involved. Structurally, RCM can be a much closer fit for this class of associative computation. Owing to the direct use of nano-scale memory array for associative computing, it can provide very high degree of parallelism, apart from eliminating the overhead due to memory read.

Associative computing of practical complexity with RCM is essentially analog in nature, as it involves evaluating the degree of correlation between inputs and the stored data. As a result, most of the designs for associative hardware using RCM's proposed in recent years, involved analog CMOS circuits for the processing task [9, 11]. Recent experiments on analog-computing with of multi-level Ag-Si memristors also employed analog operational amplifiers for current-mode processing [8]. However, application of multiple analog blocks for large scale RCM may lead to power hungry designs, due to large static power consumption of such circuits. This can eclipse the potential energy benefits of RCM for non-Boolean computing. Moreover, with technology scaling, the impact of process variations upon analog circuits becomes increasingly prominent, resulting in lower resolution for signal amplification and processing [16]. Hence, conventional analog circuits may fail to exploit the RCM technology for energy-efficient, non-Boolean computing.

The solution to this bottleneck may lie with alternate device technologies that can provide a better fit for the required non-Boolean, analog functionality, as compared to CMOS switches. Recent experiments on spin-torque devices have demonstrated high-speed switching of scaled nano-magnets with small currents [12-14]. Such magneto-metallic devices can operate at ultra low terminal voltages and can implement current-mode summation and comparison operations, at ultra low energy cost. Such current-mode spin switches or 'neurons' can be exploited in energy-efficient analog-mode computing [20-24]. In this work we present a design of RCM based associative memory using such "spin-neurons". In the proposed scheme, the spin-neurons form the core of hybrid processing elements (PE) that are employed in RCM based associative modules and achieve more than two orders of magnitude lower computation energy as compared to conventional mixed-signal (MS) CMOS circuits.

The rest of the paper is organized as follows. Section-II describes the application of RCM in non-Boolean computing and the design challenges associated with a mixed-signal implementation. The device model for spin neuron is introduced in section III. Design of analog associative memory module (AMM) using spin neurons applied to RCM modules is described in section IV. Section V discusses the performance and prospects of the proposed design scheme. Conclusions are given in section-VI.

## 2. COMPUTING WITH RESISTIVE CROSSBAR MEMORY (RCM)

Fig. 1 depicts a resistive crossbar memory. It constitutes of memristors (Ag-Si) with conductivity $g_{ij}$, interconnecting two sets of metal bars ($i^{th}$ horizontal bar and $j^{th}$ in-plane bar). Multi-level write techniques for memristors in crossbar arrays have been proposed and demonstrated in literature that can achieve precision up to 0.3% (equivalent to 8-bits) [1-2]. However, the energy-cost of the write operations may increase significantly for higher precision requirements [1-2]. In this work we have used 3% write accuracy (equivalent to 5-bits) for the memristors [8]. For a given write-precision, larger number of bits can be obtained by using parallel combination of multiple memristors to store a single analog value [4]. Note that, the class of non-Boolean pattern-matching computations, a prospective application of RCM technology, are inherently approximate and have relaxed precision constraints [11].

Memory-based pattern-matching applications generally apply some form of feature reduction technique to extract and store only the essential 'patterns' or 'features' corresponding to different data samples. The extracted patterns can be represented in the form of analog vectors that can be stored along individual columns of the RCM shown in fig.1 (note that the data is stored in the cross-point memristive element). In order to compute the correlation between an input and the stored patterns, input voltages $V_i$ (or currents $I_i$) corresponding to the input feature can be applied to the horizontal bars. Assuming the outward ends of the in-plane bars grounded, the current coming out of the $j_{th}$ in-plane bar can be visualized as the dot product of the inputs $V_i$ and the cross-bar conductance values $g_{ij}$ (fig. 1). Hence, an RCM can directly evaluate correlation between an analog input vector and a number of stored patterns. This technique can be exploited in evaluating the degree of match (DOM) between an input and the stored patterns, the best match being the pattern corresponding to the highest correlation magnitude ($\sum_i V_i g_{ij}$).

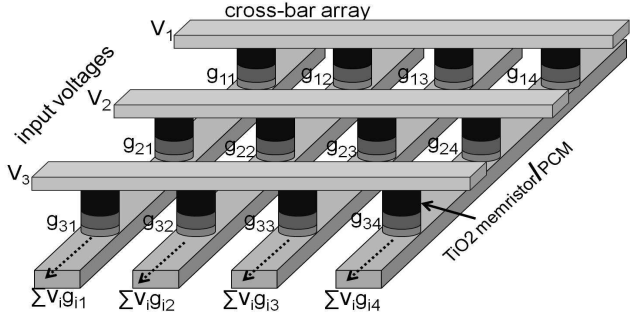

Fig. 1 A Resistive crossbar network used for evaluating correlation between inputs and stored data.

Fig. 2 depicts the feature extraction step for human face-images. In this work, we have used 10 different face-images for 40 individuals, for generating 40 stored data patterns. For an individual, each of the 10 face-images were normalized and down sized from 128x96, 8-bit pixels to 16x8, 5-bit pixels. Pixel wise average of the 10 reduced images was taken to generate 128-element (16x8), 32 level analog patterns corresponding to the 40 individual faces.

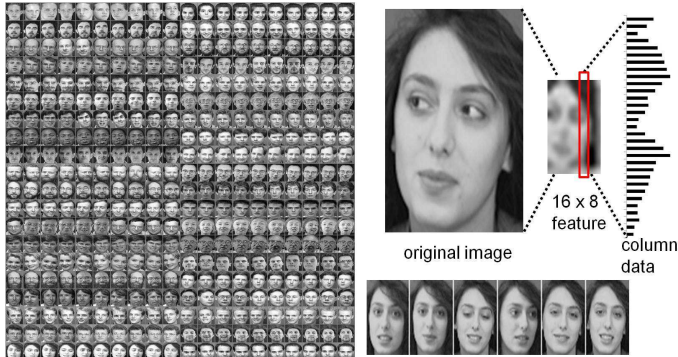

Fig. 2. 400 test images of 40 individuals (from ATT Cambridge Image Database [22]) and the feature reduction method used in this work.

The limit of image down-sizing was identified as the scaling factor below which matching accuracy for the 400 test images dropped significantly below the value achieved using the full size image (fig. 3a). For each set of downsizing factor and bit-size, current-mode correlation outputs were obtained using SPICE model of RCM.

Variations in input source as well as memristor values were incorporated to obtain realistic values for the current-outputs. For a given set of stored images, classification accuracy also depends upon the resolution of the detection unit used to determine the DOM figures for all the stored patterns. A resolution of 4% (5-bit) was chosen based on the observation that down to this value, the classification accuracy remained close to that achievable using ideal comparison (fig. 3b).

Resolving ~4% difference among the current-mode dot product requires a precision of 5-bits for the detection unit, responsible for identifying the winning pattern. Fig. 4 shows a conventional mixed-signal-CMOS solution for the detection unit. It constitutes of regulated current mirrors as the input stage, that offer low input-impedence and a near constant DC bias to the RCM. Following this, a winner-take-all (WTA) circuit receives the current inputs and determines the 'winner'. Several versions of WTA circuits have been proposed in literature, that can be classified into two broad catagories, current-conveyer WTA (CC-WTA) [18], and binary tree WTA (BT-WTA) [18]; the later being more suitable for large number of inputs [17, 18]. BT-WTA employs a binary tree of 2-input comparison stages which involve copying and propagating the larger of the two current inputs to the output (fig 4) [17].

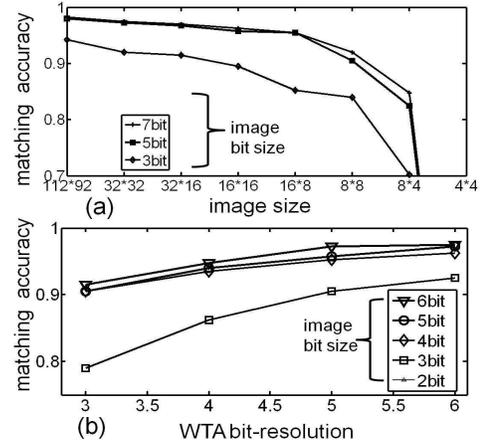

Fig. 3 (a) Training accuracy reduces with image down-sizing, (b) similar trend is obtained for reducing WTA resolution

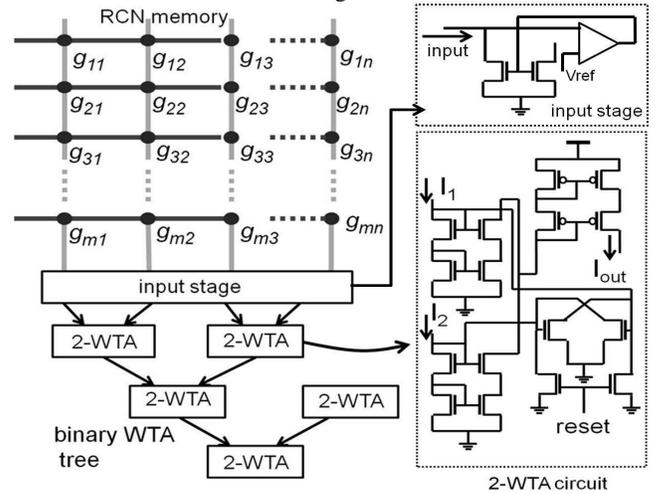

Fig. 4 A standard CMOS solution for associative memory module using binary tree winner-take-all circuit.

In general, the use of such analog WTA circuits leads to large static power consumption. In fact, the power consumption of an analog WTA unit can be several times larger than the RCM itself. Moreover, performance of such current-mirror based circuits is limited by random mismatches in the constituent transistors and other non-idealites like, channel length modulation, that introduce mismatch in different current paths [16]. In order to maintain a sufficiently high resolution, larger transistor dimensions (both length as well as width) and hence, larger cell area is needed. This is evident from some recent designs [18], that employed significantly large channel lengths for such circuits despite using relatively scaled CMOS technology. This leads to increased parasitic capacitance and hence, lower operating frequency for a given static power. Higher frequency and resolution can be achieved at the cost of increased input currents, ie., at the cost of larger power consumption [16]. Special techniques to enhance the precision of current mirrors have been proposed in literature [18], but they introduce significant overhead in terms of power consumption and area complexity. Voltage-mode processing can also be employed in RCM, however, it incurs additional overhead due to current to voltage conversion and subsequent amplifications. This results in larger mismatch, non-linearity and power consumption.

The above discussion suggests that the conventional mixed-signal CMOS design techniques may not be able to leverage the emerging nano-scale resistive memory technology for memory based computing. This motivates us to look towards alternate device technologies that can be more suitable for this purpose. In the next section we present the spin-based neuron model that can lead to efficient computing hardware based on RCM.

## 3. SPIN NEURON FOR RCM BASED COMPUTING

In this section, we describe the device operation of the spin based neuron model that is based on domain wall magnet (DWM) [12-15]. The circuit technique employed to interface the domain wall neuron (DWN) with CMOS units is also discussed.

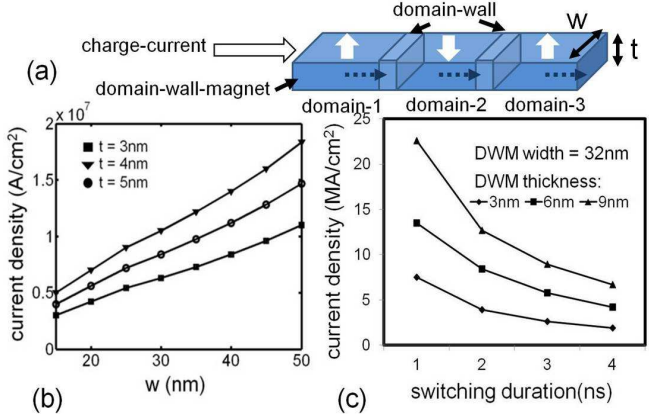

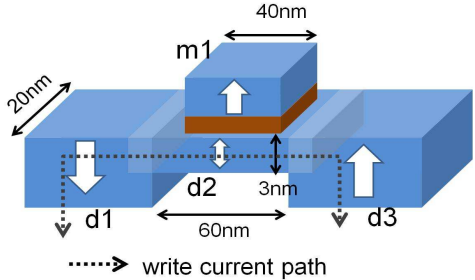

Fig. 5 (a) Domain wall magnet with three domains. (b) Scaling of DWM achieves reduction in critical switching current. (c) Smaller device dimensions achieve faster switching for a given write-current.

Fig.6 Device structure for domain wall neuron (DWN). The input current enters the device through $d_1$ and exits through $d_2$. The magnet $m_1$ associated with the MTJ is used to read the state of the free layer ($d_2$).

A domain wall magnet (DWM) constitutes of multiple *nano-magnet* domains separated by non-magnetic regions called domain-wall (DW) as shown in fig. 5a. DW can be moved along a magnetic nano-strip using current-injection. Hence, the spin polarity of the DWM strip at a given location can be switched, depending upon the polarity of its adjacent domains and the direction of current flow. Recent experiments have achieved switching current density of ~$10^6 A/cm^2$ for nano-scale DWM strips, and, a switching time of less than 1ns [12-14]. Thus, the polarity of a scaled *nano-magnet* strip of dimension $3 \times 20 \times 60 nm^3$ can be switched using a small current of ~1µA. Moreover, the current threshold as well as the switching time of DWM scales down with device dimensions (fig. 5b) [15].

The device structure for a domain-wall neuron (DWN) is shown in fig. 6. It constitutes of a thin and short ($3 \times 20 \times 60\ nm^3$) *nano-magnet* domain, $d_2$ (domain-2, the 'free-domain') connecting two anti-parallel *nano-magnet* domains of fixed polarity, $d_1$ (domain-1) and $d_3$ (domain-3). Domain-1 forms the input port, whereas, domain-3 is grounded. Spin-polarity of the free-domain ($d_2$) can be written parallel to $d_1$ or $d_3$ by injecting a small current along it from $d_1$ to $d_2$ and vice-versa [2]. Thus, the DWN can detect the polarity of the current flow at its input-node. Hence it acts as an ultra-low-voltage and compact current-comparator that can be employed in energy efficient current-mode data processing [20-24]. A non-zero current threshold for DW motion however, results in a small hysteresis in the DWN switching characteristics (fig. 7a). It is desirable to reduce the threshold to get closer to the step transfer function of an ideal comparator. Apart from device-scaling, the use lower anisotropy barrier for the magnetic-material can be effective in lowering the switching threshold for computing applications [10].

A magnetic tunnel junction (MTJ) [15], formed between a fixed polarity magnet $m_1$ and $d_2$ is used to read the state of $d_2$. The effective resistance of the MTJ is smaller when $m_1$ and $d_2$ have the same spin-polarity and vice-versa ($R_{parallel}$=~5kΩ and $R_{anti-parallel}$~15kΩ). We employ a dynamic CMOS latch shown in fig. 7b to detect the MTJ state. One of its load branches it connected to the DWN MTJ whereas the other is connected to a reference MTJ whose resistance is midway between the two resistances of the DWN MTJ. The latch effectively compares the resistance between its two load branches through transient discharge currents. Since the transient read-current flows only for a short duration, it does not disturb the state of $d_2$.

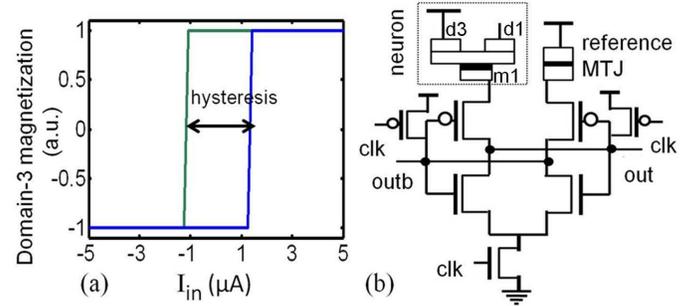

Fig. 7 (a) DWN transfer characteristics for anisotropy-energy-barrier, $E_b$ = 20KT, (b) dynamic CMOS latch used to detect DWN's state.

Note that integration of similar spin-device structure with CMOS has already been demonstrated for memory applications [15]. Notably, the energy-barrier, the threshold current as well as dimensions for a memory device needs to be larger, in order to ensure long-term stability. However, for computing applications, aggressive scaling is desirable [10]. Application of similar devices in digital logic design has also been proposed earlier [10]. In this work however, we focused on an entirely different and under-explored potential of such a spin-torque device, and showed its benefit in analog-mode non-Boolean computing.

## 4. ASSOCIATIVE MEMORY MODULE USING SPIN NEURONS AND RCM

In the following subsections, we first describe the design of RCM-based correlation unit and its interfacing with DWNs. This is followed by circuit level description of spin-CMOS hybrid-PE based on DWN that achieves the WTA functionality at ultra low energy cost.

*A. Network Design*

Fig. 8a depicts the DWNs with their input ($d_1$ terminals) connected to RCM outputs. A DC voltage, $V$, is applied to the $d_3$ terminals of all the DWNs. Owing to the small resistance of the DWN devices; this effectively biases output ends of the RCM (connected to $d_1$ terminals) to the same voltage. As described in section-2, in order to perform associative matching of an input face-image with the data stored in the RCM, the input-image is down-sized to 16x8, 5-bit pixels. Each of the 128 digital values needs to be converted into analog voltages/current levels, to be applied to the RCM-input. The low-voltage operation of DWN can be exploited to implement, compact and energy efficient current-mode DAC using binary weighted deep-triode current-source (DTCS) PMOS transistors, as shown in fig. 8a. A DC supply of $V+\Delta V$ is applied to the source terminals of the DTCS, where $\Delta V$ is ~30mV. Ignoring the parasitic resistance of the metal crossbar, the drain to source voltage of the DTCS-DAC can be approximated to $\Delta V$. The current $I_{in}(i)$, supplied by the $i_{th}$ DAC can thus be written as $\Delta V \cdot G_T(i) G_{TS}/(G_T(i) + G_{TS}))$, where $G_T(i)$ is the data dependent conductance of the $i_{th}$ DAC and $G_{TS}$ is the total conductance (of all the Ag-Si memristors) connected to a horizontal bar (dummy memristors are added for each horizontal input bar such that $G_{ST}$ is equal for all horizontal bars). As a result, the current input through a memristor connecting the $i_{th}$ input bar to the $j_{th}$ output bar (in-plane) can be written as $I(i,j)=\Delta V \cdot G_T(i) G_{ST}/(G_T(i)+G_{ST}))(G(i,j)/G_{ST})$, where, $G(i,j)$ is the programmed conductance of the memristor.

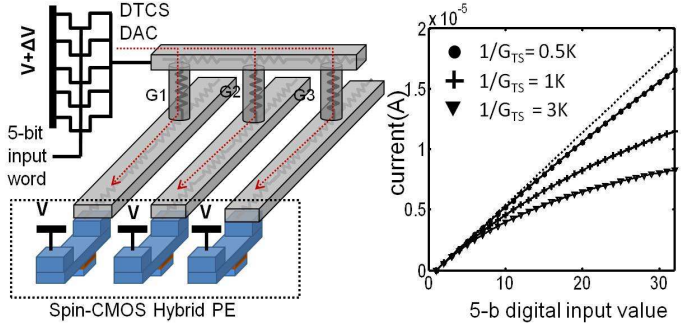

Fig. 8(a) RCM with a single DTCS input and three receiving DWN (b) non-linear characteristics of DTCN resulting due to series combination with Gs.

For accurate dot-product evaluation, the current $I(i,j)$ should be proportional to the product of $G_T(i)$ ( ie, the DTCS conductance, proportional to the input data) and $G(i,j)$. Hence, a low value of $G_{TS}$ (i.e. higher resistance values of the memristors) introduces non-linearity in the DTCS-DAC characteristics (fig. 8b). This leads to reduction in the detection margins (difference between the best and the second best match) for the current-mode dot product outputs for different input images (fig. 9a). As a result, the overall matching accuracy of the network reduces for a given WTA resolution. Ideally, choosing the lowest possible range of values for the memristor resistances (say 200Ω-6.4KΩ) would largely overcome the non-linearity (fig. 9b). However, for higher $G(i,j)$ (low resistance value for the memristors), voltage drop in the metal lines due to parasitic resistances result in corruption of the current signals, once again, leading to degradation in the detection-marging. Hence, the optimal range for the conductance values was found based on the maximum achievable read margin, as shown in fig. 9a. The design parameters like the image compression factor, data bit-width etc, discussed earlier, were therefore determined based on the simulation of RCM model, in order to ensure resolvable detection margin.

The range of current output from the DTCS-DAC needed is mainly determined by the choice of WTA resolution. If the DWNs are designed to have a threshold of ~1μA, the maximum value of the dot-product output must be greater than 32μA for a 5 bit resolution for the WTA (described later). This in turn, translates to the required range of DAC output current. For 128 element input vectors and 5-bit resolution for the WTA, the maximum value for DAC output required was found to be ~10 μA. This range of current can be obtained using different combination of DTCS sizing and the terminal voltage, $\Delta V$. For a required amount of DAC current, it is desirable to push $\Delta V$ to the minimum possible value, in order to reduce the static power consumption in the RCM. This would imply exploiting the low-voltage operation of the DWNs to the maximum possible extent. The minimum value of $\Delta V$ is limited mainly by the parasitic voltage drops that degrade the detection margin and hence the matching accuracy (fig. 9b). For this design (RCM of size 128x40) $\Delta V$ of 30mV was found to be enough to preserve the matching accuracy close to the ideal case (with no-parasitic). The proposed scheme effectively biases the RCM across a small terminal voltage ($\Delta V$), thereby ensures that the static current flow in RCM takes place across a small terminal voltage of ~30mV (between two DC supplies V and V+$\Delta V$).

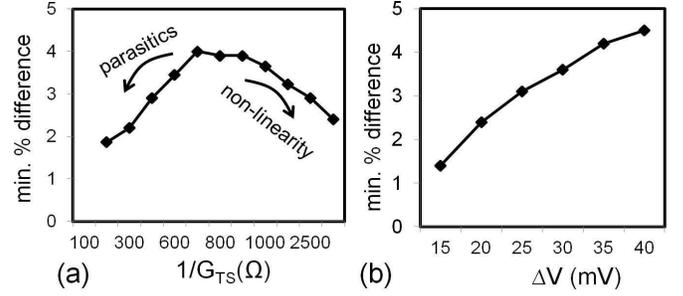

Fig. 9 (a) degradation in detection margin for a given input due to non-linearity (for low $G_{TS}$) and parasitic voltage drops (for high $G_{TS}$), (b) degradation in detection margin for the same input, for reducing $\Delta V$, due to parasitic voltage drops.

Above, we noted that the application of DWN in the RCM offers the benefit of ultra-low-voltage operation that reduces the static power-consumption resulting from current-mode, analog computing. Next we describe the design of spin-CMOS hybrid WTA that performs the winner selection task with negligible static power consumption.

### B. WTA Design

The DWN device essentially acts as a low voltage, high speed, high resolution current-mode comparator and hence can be exploited in digitizing analog current levels at ultra low energy cost [20]. The proposed WTA scheme, algorithmically depicted in fig. 10, exploits this fact and clubs a digitization step with a parallel 'winner-tracking' operation.

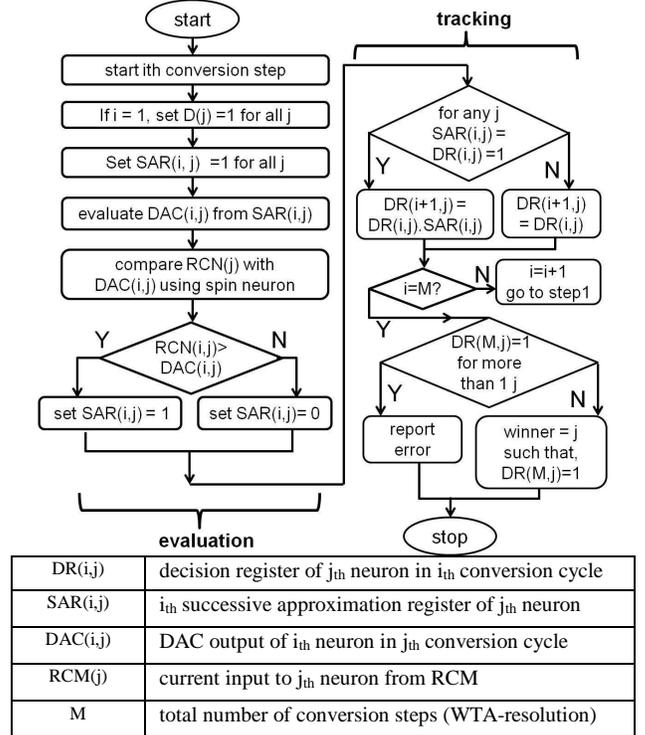

| DR(i,j) | decision register of $j_{th}$ neuron in $i_{th}$ conversion cycle |
| SAR(i,j) | $i_{th}$ successive approximation register of $j_{th}$ neuron |
| DAC(i,j) | DAC output of $i_{th}$ neuron in $j_{th}$ conversion cycle |
| RCM(j) | current input to $j_{th}$ neuron from RCM |
| M | total number of conversion steps (WTA-resolution) |

Fig. 10 WTA algorithm used in this work

The first half of the flowchart can be identified as the standard algorithm for successive approximation register (SAR) ADC [20]. The data conversion algorithm employed in an SAR-ADC can be explained as follows. To begin the conversion, the approximation register (that stores the digitization result) is initialized to the mid-scale (i.e., all but the most significant bit is set to 0). At every cycle a digital to analog converter (DAC) produces an analog level corresponding to the digital value stored in the SAR and a comparator compares it with the analog input using an analog comparator. If the comparator output is high, the current bit remains high, else it is turned low and the next lower bit is turned high. The process is repeated for all the bits. At the end of conversion, the SAR stores the digitized value corresponding to the analog input.

The circuit realization of this operation using DWN's is shown in fig. 11. Output currents of the RCM columns (in this case 40 columns storing the pattern vectors of 40 face-images) are received by individual DWN input nodes that are effectively clamped at a DC supply $V$, as described earlier. Each DWN has an associated DTCS-DAC, which is driven by the corresponding successive approximation register. The drain terminals of the DTCS transistors are a DC voltage $V-\Delta V$. In each conversion cycle, the DWN device essentially compares the RCM output and the DAC output (and hence, acts as the comparator of the SAR block). The comparison result is detected by the latch described in fig. 7, and the result is used to modify the SAR logic using the scheme described above (though pass-gate based multiplexers $P$, driven by a global controller). Note that, in the overall scheme, the component of RCM output current sunk by the DTCS in the ADC's flow through across a DC level of $2\Delta V$.

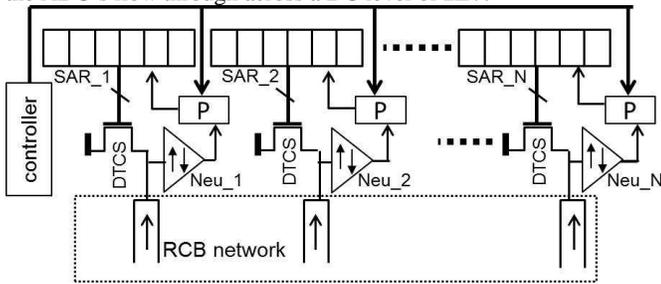

Fig. 11 Block diagram for SAR operation of the WTA circuit

The second half of the WTA algorithm operates in parallel with the first (i.e., the ADC operation). It can be explained with the help of the corresponding circuit diagram shown in fig. 12. Results of the first ADC conversion step obtained from the SAR are directly transferred to the tracking registers (*TR*) shown in the figure through the pass-gate multiplexing switch (*PGS*). Thus, at this stage, all the TR's with a high output correspond to the ADC results with MSB = '1'. Let us now, consider the second cycle operation. The detection line (*DL*) is first pre-charged to $Vdd$ and the set of discharge registers (*DR*) driving it are cleared to low output. Next, if for at least one of the SAR's with high MSB, the second MSB also evaluates to '1', the corresponding *DR* is driven high by the associated AND gate. Thus, *DL* is discharged to ground and the write of all the *TR*'s is enabled. All the *TR*'s for which both, first and second MSB's evaluated to '1', stay high, but the rest are set to low. In simple terms, if at least one of the SAR's (5-bit) evaluated to '11000' in the second conversion cycle, the *DL* is discharged and all the *TR*'s with SAR value '11000' stay high, while those with SAR value '10000' are set to low. In case all SAR's evaluated to '10000' in the second cycle, no change is made to the *TR* values. Thus, at the end of conversion cycle, if only one of the *TR*'s remains high, it is identified as the winner and the corresponding SAR value is effectively the degree of match (DOM). In case a random image is input to the hardware, the proposed scheme will still identify the 'winning' pattern. But if the DOM is lower than a predetermined threshold, the winner is discarded, implying that the input image does not belong to the stored data set.

The winner-tracking circuitry described above is fully digital and does not consume any static power. Moreover, owing to the global digital control, it is easily scalable with number of input as well as required bit precision. The overall power consumption in the proposed design is drastically reduced as compared to a MS-CMOS realization (described in section-2), due to two main reasons; firstly, the power consumption in the RCM itself is significantly lowered due to low-voltage operation, and secondly, the fully digital WTA scheme avoids any additional static-power consumption. Note that the proposed WTA scheme implemented in MS-CMOS would result in large power consumption, resulting from conventional ADC's. The low-voltage current-mode switching characteristics of DWN however, provides a compact and ultra-low-power digitization technique [20, 22]. The performance and prospects of the proposed design are discussed in some more detail in the next section.

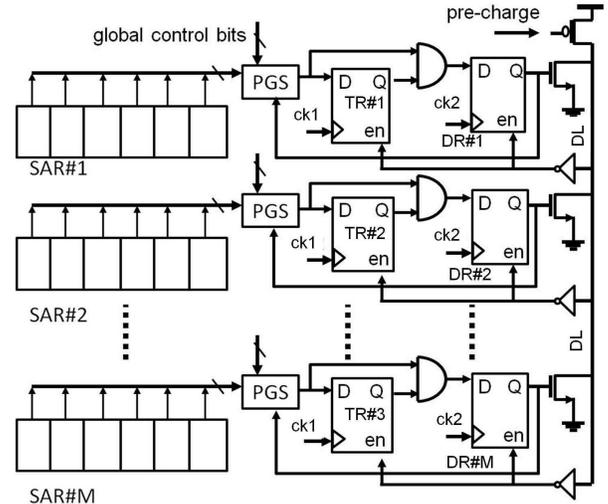

Fig.12 Circuit operation for the tracking part of the WTA algorithm

## 5. PERFORMANCE AND PROPECTS

In order to compare the performance of the proposed design with state of the art mixed signal (MS) CMOS design, we simulated two different CMOS BT-WTA topologies proposed in [17] and [18] respectively, using 45nm CMOS technology models. The first design is the standard BT-WTA, whereas, the second is a recently proposed modification. We also simulated a 45nm digital CMOS design that employed multiply and accumulate operations for evaluating the correlation between the 5-bit 128 element digital templates and input features of the same size.

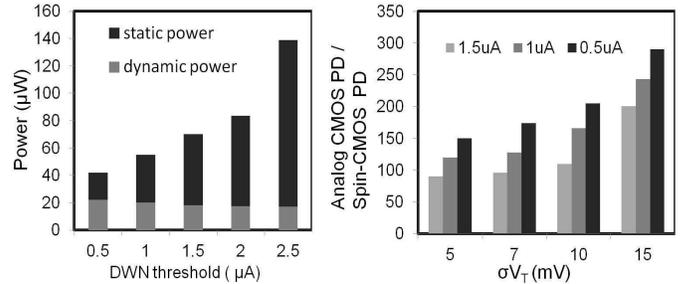

Fig. 13 (a) Power consumption of the proposed design with its static and dynamic components, for different values of DWN threshold, (b) ratio of power-delay (PD) product of MS-CMOS and the proposed design for increasing transistor variations.

Simulations for MS-CMOS designs show that the power consumption for the WTA unit dominates the total power. On the other hand, for the proposed scheme, there is negligible static power consumption in the WTA operation. However, since, the static power consumption in RCM is also significantly lowered, it becomes comparable to the dynamic switching power in the WTA. This is evident from the trend shown in fig. 13a. It also shows that the static power consumption in the DWN-based design can be significantly reduced by lowering the DWN switching threshold further. However, the dynamic power remains almost constant and starts to dominate for reduced DWN thresholds.

Plot in fig. 13b shows the impact of transistor process variations upon MS-CMOS designs. The power-delay products are plotted for a WTA resolution of 4%. Note that in the proposed WTA, the impact of transistor-variations in the DTCS-DAC is limited to just a single step, whereas, the MS-CMOS circuits suffer more due to the cumulative effect of multiple transistors in the signal path. As discussed in section-2, with larger variations, the accuracy and resolution of MS-CMOS circuits like, current-mirrors decreases steeply, necessitating the use of larger devices, which impairs the circuit performance.

Table-1 compares the proposed spin-CMOS design with MS-CMOS designs in [17] and [18], and with the 45 nm digital CMOS design. The results shown are for $\sigma V_T$ =5mV for minimum sized transistors, which is a near ideal case for MS-CMOS circuits. Results for three different WTA resolutions are given which show similar energy benefits of the proposed scheme, even for smaller WTA resolution. For analog designs, lower resolution constraints allows smaller transistors and hence better performance. Power consumption for the DWN based design, also reduces with resolution. Lower WTA resolution allows smaller DAC currents, resulting in reduced static power and lower switched capacitance for the smaller WTA blocks, leading to reduced dynamic power.

Table-1 Performance

|  |  | spin-CMOS PE | [ 18 ] | [ 17 ] | 45nm Digital CMOS |
|---|---|---|---|---|---|
| Power | 5-bit | 65µW | 5.5mW | 8mW | 4mW |
|  | 4-bit | 45µW | 2.9mW | 5.0mW | 2.8mW |
|  | 3-bit | 32µW | 2.3mW | 3.2mW | 1.2mW |
| Frequency |  | 100 MHz | 50MHz | 50MHz | 2.5MHz |
| Energy | 5-bit | 1 | 160 | 215 | 2460 |
|  | 4-bit | 1 | 140 | 221 | 2300 |
|  | 3-bit | 1 | 155 | 210 | 1100 |

Most interestingly, results for comparison with 45nm digital hardware shows ~1000x lower computing energy for the proposed design. Note that, this comparison does not include the overhead due to memory read in the digital design. As discussed earlier, digital hardware in general prove inefficient for the class of computation considered in this work. Another important point to be noted is that, the use of MS-CMOS circuits in RCM barely perform ~10x better than the digital implementation and hence achieve far less energy efficiency as compared to the proposed design. Thus, ultra-low energy analog-computing using spin-neurons can significantly enhance the prospect of RCM technology for computational hardware. The basic associative-module discussed in this work can be extended to a more generic architecture. For instance, very large number of images can be grouped into smaller clusters [25], that can be hierarchically stored in the multiple RCM modules. Individual patterns of larger dimensions can also be parititioned and stored in modular RCM-blocks. The proposed design scheme can be applicable to a wide class of non-Boolean computing architectures that also include different categories of neural networks [20]. For instance, the spin-RCMbased correlation modules presented in this work can provide energy efficient hardware solution to convolutional neural networks that are attractive for cognitive computing tasks, but involve very high computational cost.

A self explanatory pictorial depiction of the simulation-framework used in this work is given in fig.14. We used micromagnetic simulation model for DWN that was calibrated with experimental data on DWMs. Behavioral model based on statistical characteristics of the device were used in SPICE simulation to assess the system level functionality. Some important design parameters used are listed in table-2.

## 6. CONCLUSION
Emerging RCM technology holds great potentials for non-Boolean computing hardware. However, conventional mixed signal CMOS circuits may fail to leverage its benefits due to their large power consumption and poor scalability. We showed that the critical analog functionality needed in RCM based computing tasks can be provided by magneto-metallic spin-neurons at ultra low energy cost. The resulting design can achieve more than three orders of magnitude lower energy cost as compared to a dedicated digital hardware. The use of spin-torque neurons can therefore boost the prospects of RCM as a computation tool.

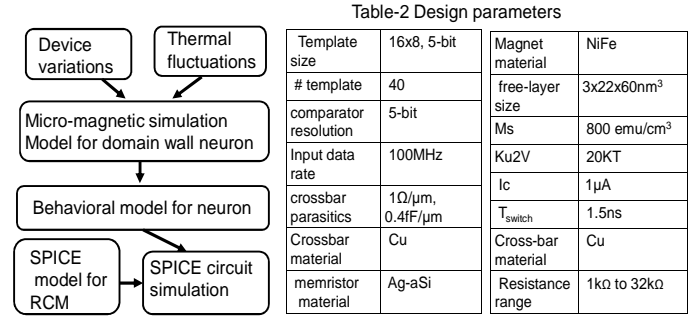

Table-2 Design parameters

| Template size | 16x8, 5-bit | Magnet material | NiFe |
|---|---|---|---|
| # template | 40 | free-layer size | $3x22x60nm^3$ |
| comparator resolution | 5-bit | Ms | 800 emu/$cm^3$ |
| Input data rate | 100MHz | Ku2V | 20KT |
| crossbar parasitics | 1Ω/µm, 0.4fF/µm | Ic | 1µA |
| Crossbar material | Cu | $T_{switch}$ | 1.5ns |
| memristor material | Ag-aSi | Cross-bar material | Cu |
|  |  | Resistance range | 1kΩ to 32kΩ |

Fig. 14 Simulation framework

**ACKNOWLEDGEMENT**
This work was supported by STARnet, a Semiconductor Research Corporation program sponsored by MARCO and DARPA.